\documentclass[twocolumn,aps,prl,superscriptaddress,showpacs]{revtex4}
\usepackage{hyperref}

\usepackage{graphicx}

\usepackage{amsmath}
\usepackage{epsfig}

\begin{document}
 \title{Demonstration of near-Optimal Discrimination of Optical Coherent States}

\author{Christoffer Wittmann}
\affiliation{Institut f\"{u}r Optik, Information und Photonik, Max-Planck Forschungsgruppe, Universit\"{a}t Erlangen-N\"{u}rnberg, G\"{u}nther-Scharowsky str. 1, 91058, Erlangen, Germany}

\author{Masahiro Takeoka}
\affiliation{National Institute of Information and Communications Technology, 4-2-1 Nukui-kitamachi, Koganei, Tokyo 184-8795, Japan}
\affiliation{CREST, Japan Science and Technology Agency, 1-9-9 Yaesu, Chuoh-ku, Tokyo 103-0028, Japan} 

\author{Kati\'uscia N. Cassemiro}
\affiliation{Instituto de F\`isica, Universidade de S\~ao Paulo, Caixa Postal 66318,
S\~ao Paulo, SP, Brazil, 05315-970}

\author{Masahide Sasaki}
\affiliation{National Institute of Information and Communications Technology, 4-2-1 Nukui-kitamachi, Koganei, Tokyo 184-8795, Japan}
\affiliation{CREST, Japan Science and Technology Agency, 1-9-9 Yaesu, Chuoh-ku, Tokyo 103-0028, Japan} 

\author{Gerd Leuchs}
\affiliation{Institut f\"{u}r Optik, Information und Photonik, Max-Planck Forschungsgruppe, Universit\"{a}t Erlangen-N\"{u}rnberg, G\"{u}nther-Scharowsky str. 1, 91058, Erlangen, Germany}

\author{Ulrik L. Andersen}
\affiliation{Department of Physics, Technical University of Denmark, Building 309, 2800 Kgs. Lyngby, Denmark}
\affiliation{Institut f\"{u}r Optik, Information und Photonik, Max-Planck Forschungsgruppe, Universit\"{a}t Erlangen-N\"{u}rnberg, G\"{u}nther-Scharowsky str. 1, 91058, Erlangen, Germany}

\date{\today}

\begin{abstract}
The optimal discrimination of non-orthogonal quantum states with minimum error probability is a fundamental task in quantum measurement theory as well as an important primitive in optical communication. In this work, we propose and experimentally realize a new and simple quantum measurement strategy capable of discriminating two coherent states with smaller error probabilities than can be obtained using the standard measurement devices; the Kennedy receiver and the homodyne receiver. 
\end{abstract}


\maketitle 
One of the most profound consequences of quantum mechanics is that it is impossible to construct a measurement device that perfectly can discriminate between non-orthogonal, that is overlapping, quantum states~\cite{helstromBook}. Suppose for example one is given one of two a priori known coherent states (possibly representing binary information), then there is no physical measurement that with certainty can identify which state was at hand due to the intrinsic non-orthogonality of coherent states. Since perfect discrimination without any ambiguity is impossible the canonical task is to construct a measurement apparatus that maximizes the information gained or minimizes the errors in the measurement. Such a task has received a lot of attention due to its intricate connection with fundamental quantum mechanics and due to its central role in optical communication~\cite{Giovannetti}.  
  
The impossibility of discriminating non-orthogonal quantum states is on the one hand the engine of quantum key distribution~\cite{Bennett84}, but on the other hand also a hindrance for efficient classical communication. Non-orthogonality prevents an eavesdropper from acquiring information without disturbing the state and it leads to unwanted errors in classical communication. The latter is in particular a problem in amplification-free transmission media such as is the case for deep space communication where the receiver station detects low amplitude (thus largely overlapping) coherent states. In both communication scenarios, however, it is desirable to perform the discrimination with minimum error in order to obtain the larger mutual information between sender and receiver. 

The minimum error in distinguishing between two non-orthogonal states 
was found in a pioneering work by Helstrom~\cite{helstrom,helstromBook}. 
Particularly for two weak coherent states, the physical realization of 
the measurement was later suggested by Dolinar~\cite{dolinar} and 
a proof of principle experiment has recently been reported~\cite{cook07.nat}. 
Dolinar's idea is an extension of a much simpler scheme proposed earlier
by Kennedy~\cite{kennedy} and achieving near optimal performance. For 
very weak coherent state, a simple homodyne receiver is also near optimal. 
In this letter we propose and experimentally realize a new quantum measurement that detects binary optical coherent states with fewer errors than the homodyne and the Kennedy receiver for all amplitudes of the coherent states. Although the scheme is not capable of achieving the Helstrom bound the implementation discriminates binary coherent states with an error probability lower than the optimal Gaussian receiver, namely the homodyne receiver.  

Consider the binary alphabet comprising two pure and phase shifted coherent states $\{|\alpha\rangle,|-\alpha\rangle\}$ occuring with the a priori probabilities $p_1$ and $p_2$. The task of the receiver is to certify with minimum error probability whether the state was prepared in $|\alpha\rangle$ or $|-\alpha\rangle$ using a measurement described by the two-component positive operator-valued measure (POVM) $\hat\Pi_i, i=1,2$ where $\hat\Pi_i > 0$ and $\hat\Pi_1+\hat\Pi_2=\hat I$. The average error probability is given by 
\begin{equation}
p_E=p_1\langle\alpha|\Pi_2|\alpha\rangle+p_2\langle -\alpha|\Pi_1|-\alpha\rangle
\label{errorrate}
\end{equation}
where $\langle\alpha|\Pi_2|\alpha\rangle$ ($\langle -\alpha|\Pi_1|-\alpha\rangle$) represents the error probability of mistakenly guessing $|-\alpha\rangle$ $(|\alpha\rangle)$ when $|\alpha\rangle$ $(|-\alpha\rangle)$ was prepared. In this paper we assume $\alpha$ is real and the two states to be prepared with the same probabilities: $p_1=p_2=1/2$.

According to the laws of nature, the smallest error in discriminating the two coherent states is 
\begin{equation}
p_{M}=\frac{1}{2}\left(1-\sqrt{1-e^{-4|\alpha|^2}}\right)  
\label{pM}
\end{equation} 
which is referred to as the Helstrom bound~\cite{helstrom,helstromBook}. This minimum error probability can in principle be achieved by using linear optics, a photon counter and ultra-fast feedback~\cite{dolinar} (or feedforward~\cite{takeoka05.pra}), or alternatively, using a highly nonlinear unitary operation~\cite{sasaki96.pra}. Although a very recent proof-of-principle experiment has been made~\cite{cook07.nat}, its implementation possess a high level of complexity. 
Another much simpler and near optimal approach is the Kennedy receiver, 
in which the states $\{|\alpha\rangle,|-\alpha\rangle\}$ are unconditionally displaced to $\{|2\alpha\rangle ,|0\rangle\}$, and the resulting states are detected using direct photon counting represented by the elements $\hat\Pi_1=\hat I -|0\rangle\langle 0|$ and $\hat\Pi_2=|0\rangle\langle 0|$.
The average error probability is given by 
\begin{equation}
p_K=\frac{1}{2}\langle2\alpha| \hat\Pi_2 |2\alpha\rangle
=\frac{1}{2}e^{-4|\alpha|^2} .
\label{pK}
\end{equation} 
The simplest scheme for discriminating phase shifted coherent states is, however, homodyne detection: The local oscillator is set to enable a measurement along the excitation of the coherent states, and positive measurement outcomes identifies $|\alpha\rangle$ whereas negative outcomes identifies $|-\alpha\rangle$. The POVMs are $\hat\Pi_1=\int_0^\infty|x\rangle\langle x|dx$ and $\hat\Pi_2=\hat I-\hat\Pi_1$, and the error probability is
\begin{equation}
p_H=\frac{1}{2}\left(1-{\rm erf}\left(\sqrt{2}|\alpha |\right)\right)
\label{pH}
\end{equation}
Interestingly, it has recently been proven in ref.~\cite{takeoka08} that the simple homodyne receiver is optimal within all possible Gaussian measurements.

\begin{figure}[h]
\begin{tabular}{l}
(a)\hspace{3.5cm}(b) \\ [-0.3cm]
            \centerline{\includegraphics[width=8cm,keepaspectratio]{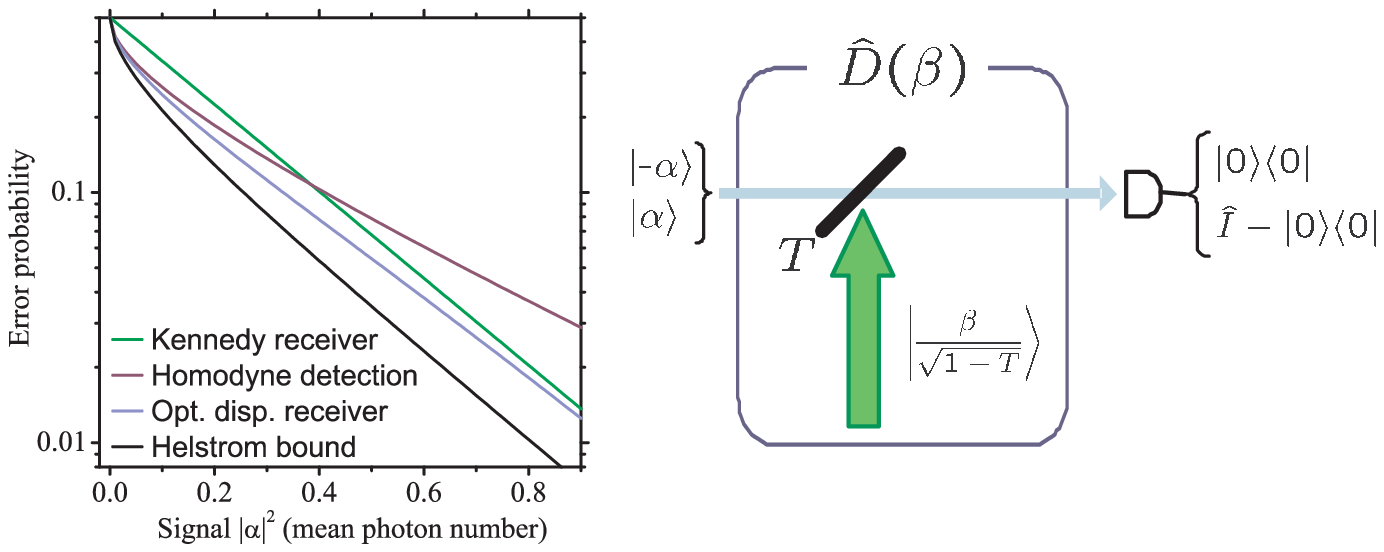}} \\ [+0.3cm]
        (c) \\ [-0.3cm]
        \centerline{\includegraphics[width=8cm,keepaspectratio]{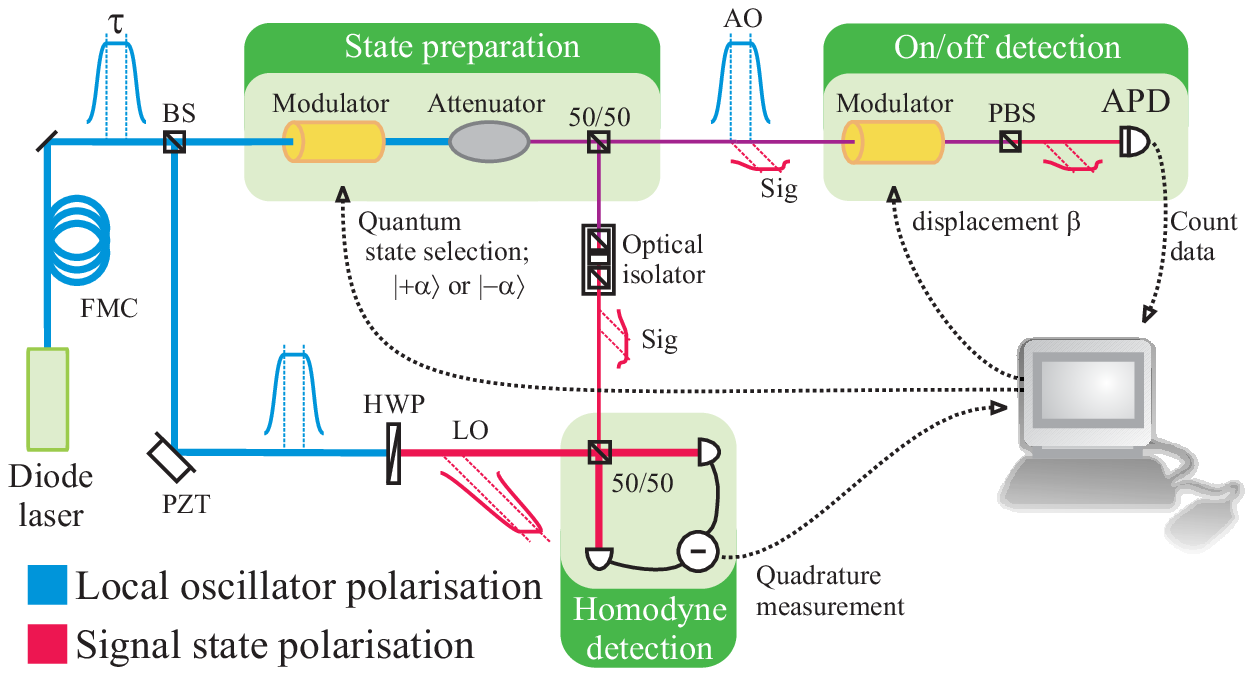}} \\
        \end{tabular}
    \caption{\label{scheme} (a) Comparison the error probabilities of the four ideal detection schemes (b) Schematic of the proposed receiver. Note that, in the limit of 
$T \to 1$, the interference exactly acts as a displacement operation 
$\hat D (\beta)$. (c) Simplified scheme of the experiment.}
\end{figure}

The three error probabilities (\ref{pM}), (\ref{pK}) and (\ref{pH}) are shown in Fig.~\ref{scheme}(a) by the solid black, green and violett curves respectively. It is evident from the figure that for most values of the coherent state amplitude the Kennedy receiver is better than the homodyne receiver. However, at very low amplitudes which is the case for quantum communication and deep space communication, the simple homodyne receiver outperforms the Kennedy receiver. 

In the following we propose a new simple receiver which outperforms the homodyne as well as the Kennedy receiver for all amplitudes. The new receiver is a slight modification of the Kennedy receiver as sketched in Fig.\ref{scheme}(b). Instead of displacing the states $|\alpha\rangle$ and $|-\alpha\rangle$ by $\alpha$, as done in Kennedy's approach, in our new receiver the states are displaced by an optimised value $\beta$ so as to minimize the error probability. In Kennedy's scheme only the error probability of detecting $|-\alpha\rangle$ by $\hat\Pi_1$ (corresponding to the second term in (\ref{errorrate})) is minimized. However, the sum of the two probabilities in (\ref{errorrate}) is not necessarily minimized; the first term (corresponding to the probability of detecting $|\alpha\rangle$ with $\hat\Pi_2$) is getting smaller the larger the displacement. Thus there exist a trade-off between these two error components, and in our new receiver we seek to minimize the sum of the two probabilities with respect to the displacement $\beta$.
The displacement is implemented by interfering the signal state with an auxiliary coherent state oscillator, $|\beta(1-T)^{-\frac{1}{2}}\rangle$, on a very asymmetric beam splitter (with transmittance $T \sim 1$) as shown in Fig.\ref{scheme}(b). The resulting displaced state is directed to a photon counter described by the projectors $\hat\Pi_1=\hat I -|0\rangle\langle 0|$ and $\hat\Pi_2=|0\rangle\langle 0|$.

After passing the beam splitter, the signal states $|\pm\alpha\rangle$ 
are transformed as $|\pm\alpha\rangle \to |\pm\sqrt{T}\alpha+\beta\rangle$ 
and the average error probability is given by 
\begin{equation}
p_\beta = \frac{1}{2} - e^{-\nu-\eta(T|\alpha|^2 + |\beta|^2)} 
\sinh \left( 2\eta\xi\sqrt{T}\alpha \beta \right) , 
\label{pBeta}
\end{equation}
where some practical imperfection parameters are included: 
$\eta$ and $\nu$ are the quantum efficiency and dark count rate of 
the photon counter and $\xi$ is the visibility of the interference at the asymmetric beam splitter. 

The optimal displacement $\beta$ is derived from the derivative $d p_\beta/d \beta = 0$, which gives the following optimal condition:
\begin{equation}
\xi \sqrt{T} \alpha = 
\beta \tanh \left(2\eta \xi \sqrt{T}\alpha\beta\right).
\label{optimalBeta}
\end{equation}
The ideal error probability for such optimized displacement receiver ($\eta =1$, $\nu =0$ and $\xi =1$) is plotted in Fig.~\ref{scheme}(a) and we see that its performance surpasses those of the homodyne and Kennedy receivers. {\itshape Noteworthy, our simple receiver outperforms any Gaussian measurement approach}. A detailed theoretical account of this new receiver is given elsewhere~\cite{takeoka08}. 

\begin{figure}
\begin{tabular}{l}
(a)\\[-0.3cm]
\centerline{\includegraphics[width=6.5cm]{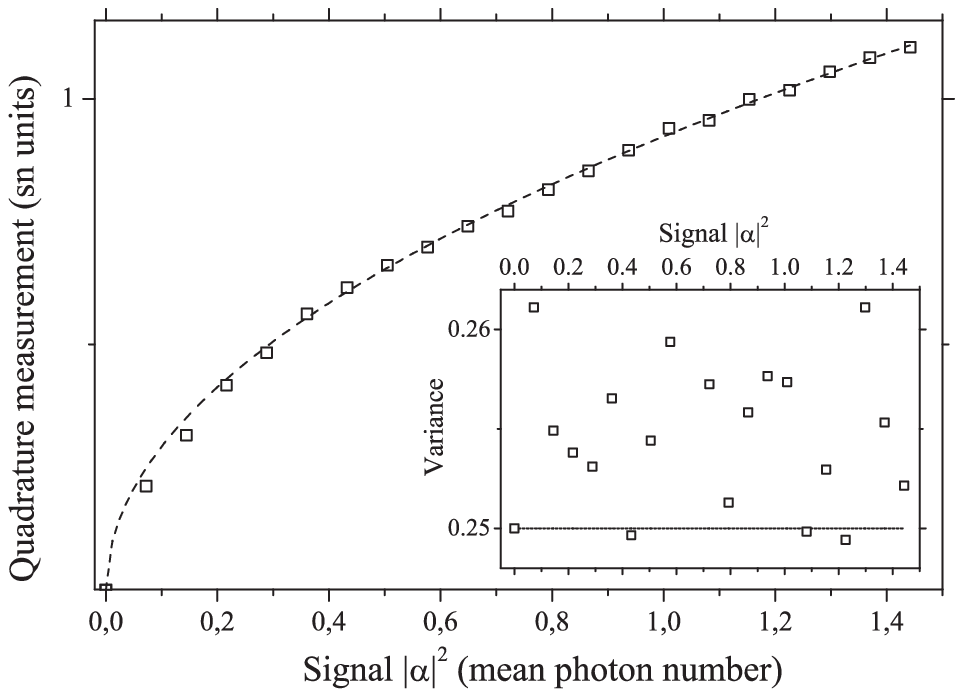} } \\ [+0.5cm]
(b)\\[-0.5cm]
\centerline{\includegraphics[width=6.5cm]{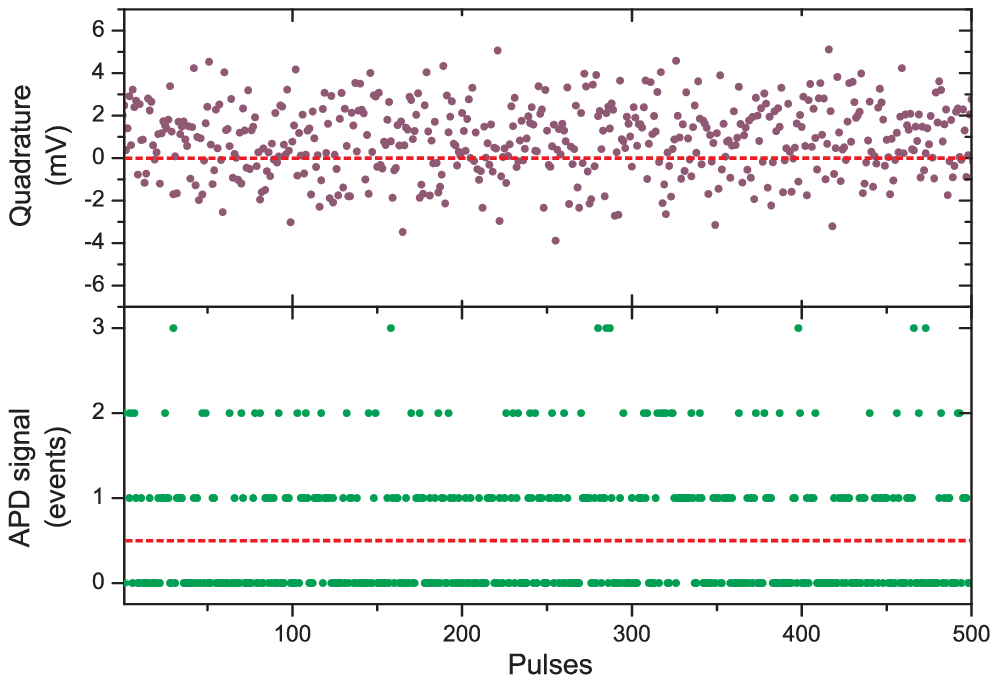}} \\ 
\end{tabular}
\caption{(a) The figure shows the mean value of quadrature measurements normalized by the shot noise, i.e. $\langle (\Delta \hat X_\theta)^2\rangle=1/4$. Data is compared to the theoretical prediction (dashed line). The amount of excess noise originating from imperfect state preparation is shown in the inset. (b) Raw measurement data for homodyne and APD detection. In the upper trace, the result of single shot quadrature measurements are shown. The lower trace shows the number of click events in the APD per pulse. The thresholds used for discrimination are shown as red dashed lines, which lie at $U=0\mathrm{mV}$ (for the quadrature measurement) and at $n=0.5$ (for APD measurement). }
\label{raw}  
\end{figure}

We proceed with a description of the experimental setup, which is shown in Fig.~\ref{scheme}(c). It consists of a preparation stage and two different receiver stages; our new receiver (which can also be made to function as a Kennedy receiver) and a homodyne receiver. Our source is a grating stabilized CW diode laser at $810\:\mathrm{nm}$ with a linewidth of $1\:\mathrm{MHz}$. After passing a fiber mode cleaner (FMC), the linearly polarized beam is split asymmetrically in two parts to serve as a local oscillator of the homodyne receiver (LO) and an auxiliary oscillator (AO) for state preparation and displacement in the new receiver scheme. The signal state (Sig) is generated in a polarisation mode orthogonal to the auxiliary mode using an electro-optical modulator: The field amplitude of the auxiliary mode is coherently transfered into the signal polarisation and the excitation is controlled by the input voltage of the modulator. Note that the auxiliary oscillator remains in the polarisation mode orthogonal to the signal mode thus propagating  along with the signal. After splitting the signal on a 50/50 beam splitter, two identical signal states (either $|\alpha\rangle^{\otimes 2}$ or $|-\alpha\rangle^{\otimes 2}$) are produced and subsequently directed to the two detection schemes. 

At the homodyne receiver the signal interferes with the local oscillator, the two resulting outputs are detected and the difference current is produced. This yields an integrated quadrature value for each signal pulse. The overall quantum efficiency of the homodyne receiver amounts to $\eta_{\mathrm{homodyne}}=85.8\%$; the interference contrast to the local oscillator is $96.6\pm0.1\%$ and the PIN-diode quantum efficiency is $92\pm3\%$. The electronic noise level is more than $23\:\mathrm{dB}$ below the shot noise level for a local oscillator power of $5\mbox{mW}$. 

The optimised displacement receiver is composed of a displacement operation and a fiber coupled avalanche photo diode (APD) operating in gated mode thus yielding an electronic pulse when a photon or more impinges onto it (thus implementing the POVMs $\hat\Pi_1=\hat I -|0\rangle\langle 0|$ and $\hat\Pi_2=|0\rangle\langle 0|$). In contrast to the displacement operation depicted in Fig.~\ref{scheme}(b) where two spatially separated modes interfere on a beam splitter, in our setup the two modes (the auxiliary and the signal modes) are in the same spatial mode but have different polarisation modes (Fig.~\ref{scheme}(b)). The interference (and thus the displacement) is therefore controlled by a modulator and a polarizing beam splitter. This method facilitates the displacement operation and yields a very high interference contrast of 99.6\%. The detection efficiency of the scheme is estimated to $\eta_{\mathrm{on/off}}=55\%$, including the transmission coefficient of the modulator, the polarisation optics and the fiber of $89.1\%$ as well as the quantum efficiency of the APD of $63\pm3\%$. The latter efficiency was estimated by the APD click statistic for an input coherent state that was calibrated by the homodyne receiver. An optical isolator is used between the two detection schemes to prevent back scattering of the LO to the APD.

The signal states are generated in time windows of $\tau=800\:\mathrm{ns}$ with a repetition rate of $100\:\mathrm{kHz}$. Several vacuum states and signal amplitudes are tested in a repeated pulse sequence. First, we carefully characterised the prepared input signal. In Fig.~\ref{raw}(a), mean values of quadrature measurements are shown for signal state ensembles with linearly increasing mean photon number in the signal. The true signal amplitude is inferred using the known quantum efficiency of the homodyne detection and the shot noise of the vacuum states. The variance is also calculated (shown in the inset) and it indicates that the prepared states are practically shot noise limited corresponding to a variance of 1/4; an average excess noise of only 0.005 shot noise units is observed.  

We proceed by describing the principles of the discrimination task. A PC acquires simultaneously the homodyne and APD detection outcomes in a pulse sequence. An example of measurements of such a sequence for $\alpha =0.35$ is shown in Fig.~\ref{raw}(b). The outcomes of the homodyne receiver are continuous quadrature values and if the value is positive we guess $|\alpha\rangle$ (which is a correct guess) and if the value is negative we guess $|-\alpha\rangle$ (which gives an error). The data from the new receiver is also shown, but here the outcomes are integers, and we use the hypothesis that if the outcome is larger than zero, we guess $|\alpha \rangle$ otherwise $|-\alpha\rangle$. The error probability is therefore found by adding up all the false detections and relate it to the total number of pulses in a sequence.

\begin{figure}[h]
\begin{tabular}{l}
(a)\\[-0.1cm]
\centerline{\includegraphics[width=7.4cm]{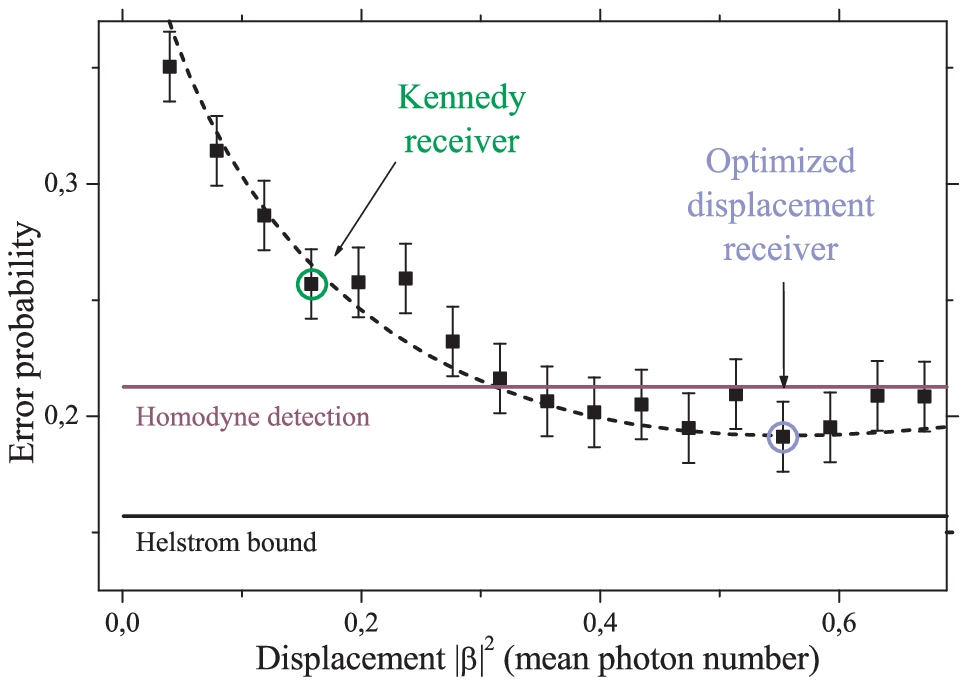}} \\ [0.5cm]
(b)\\ [-0.5cm]
\centerline{\includegraphics[width=7.5cm]{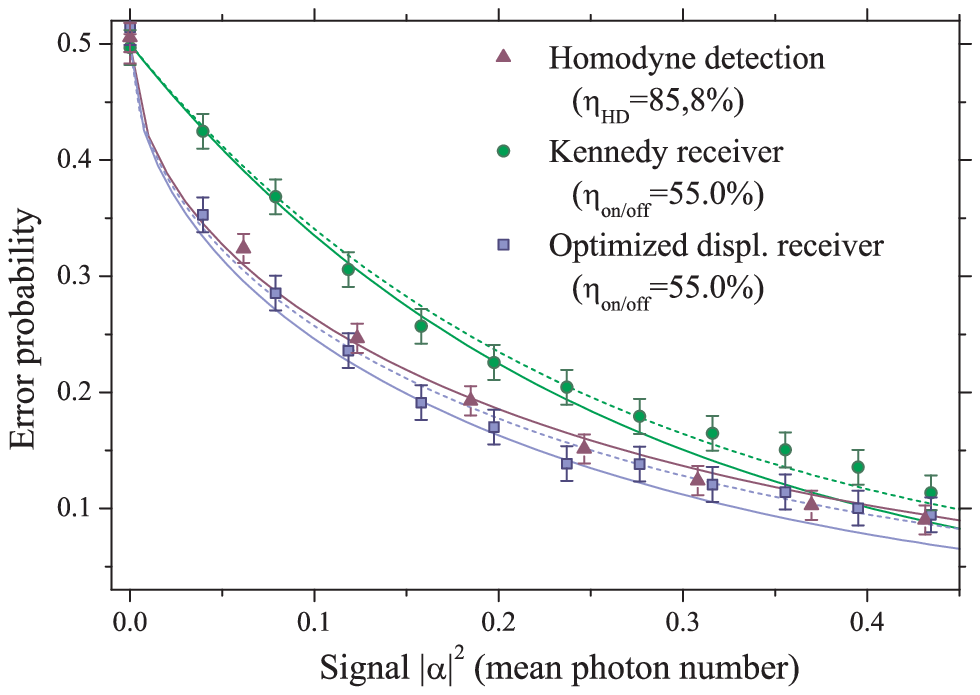}} \\ [0.5cm]
\end{tabular}
\caption{(a) Effect of displacement $\beta$ on error probability $p_\beta$ for a given signal amplitude $|\alpha|^2=0.16$ (corrected for quantum efficiency). Error rates for the Kennedy receiver (green) and optimal displacement receiver (blue) are marked. Experimental data is compared to a perfect model, without experimental imperfections (dashed line), limits for optimal discrimination (black) and Homodyne detection (violett). (b) Error rates for the detection schemes versus signal amplitude (corrected for quantum efficiency). All data points were obtained from at least 1400 measurement trajectories. Error bars reflect the standard deviations of repeated measurements, which are larger than the statistical errors. Experimental data is compared to ideal receivers (solid lines) and on/off detection schemes with experimental imperfections (dashed lines). In the demonstration of the optimal displacement receiver and the comparison of the detection schemes, we correct the signal amplitude for the quantum efficiency of the receivers. This can be justified, because detection efficiency ($\eta_{\mathrm{on/off}}$ and $\eta_{\mathrm{HD}}$) factors out in the comparison between the shot noise error and quantum limits corresponding to ideal ($\eta=1$) detection~\cite{geremia04.pra}.}
\label{optweakLO}  
\end{figure}

First we investigate the effect of the displacement $\beta$ on the error rate $p_\beta$. Note that the setup performs the displacement by tailoring the transmission coefficient, $T$, while keeping the AO amplitude, $\gamma=\beta(1-T)^{-\frac{1}{2}}$, constant.
The optimal $T$ for given $\gamma$ is  derived from the derivative $d p_\beta/d T = 0$
which gives the optimal condition 
\begin{equation}
\frac{\xi \alpha\gamma(1-2T)}{(|\alpha|^2-|\gamma|^2)\sqrt{T(1-T)}} = 
\tanh \left(2\eta \xi \sqrt{T(1-T)}\alpha\gamma\right).
\label{optimalBetaT}
\end{equation}
By varying $\gamma$, we find, that for a wide range of signal amplitudes, a minimal average error probability is achieved, if $|\gamma|^2=24.7$. This optimal AO power is chosen in the following measurements. (In practice there will always be non-zero information leaking out from the displacement operation due to the finite size of the local oscillator. By measuring this information and combining the results of the two outputs, the discrimination error is slightly lowered.)

We record the error rate for various displacements for a signal amplitude of $|\alpha|^2=0.16$. The results are shown in Fig.~\ref{optweakLO}(a), and we clearly see that the displacement has a big effect on the error rate. The data also shows, that the performance of the Kennedy receiver (corresponding to $|\alpha|^2=|\beta|^2=0.16$ and marked in the figure) is surpassed by using a larger displacement. The dashed line represents the theoretical preciction for the optimal model, and the two solid lines are the error rates associated with ideal homodyne detection (experimental comparison follows later) and the ideal, hypothetical Helstrom bound. Note that the error rate of our new receiver is lower than that of an optimal homodyne receiver even including the error bars. 

In Fig.~\ref{optweakLO}(b) we present the measured error probability for three different receivers; the homodyne receiver, the Kennedy receiver and the new optimised displacement receiver and compare the measurements with the theory. Note, that the theoretical predictions are within the error bars of the experimental data points. The graph verifies that by optimising the displacement, the performance of the Kennedy receiver can be drastically increased. The performance of the homodyne receiver and the optimal displacement receiver are comparable. However, the data set strongly indicates that the new receiver performs better than the homodyne receiver.  

In conclusion, we have experimentally demonstrated a substantial reduction in the error rate in discriminating binary optical coherent states by using a new detection approach. Remarkably, the scheme beats the optimal Gaussian approach (which is homodyning) without using optical nonlinearities or complicated feedback. Such a simple receiver may find a wide range of applications in classical as well as quantum communication.

This work has been supported by the EU project SECOQC, Lundbeckfonden(R13-A1274), MEXT Grant-in-Aid 
for Young Scientists (B) 19740253 and the Funda\c{c}\~ao de Amparo \`a Pesquisa do Estado de S\~ao Paulo (FAPESP).

\end{document}